# Approaching Landauer's Bound In A Spin-Encoded Quantum Computer

**Frank Z. Wang[1]**
[1]Division of Computing, Engineering & Mathematics Sciences, University of Kent, Canterbury, CT2 7NF, the UK
Correspondence to f.z.wang@kent.ac.uk

**ABSTRACT** It is commonly recognized that Landauer's bound holds in (irreversible) quantum operations. In this study, we verified this bound by extracting a single spin from a spin-spin magnetic interaction experiment to demonstrate that Landauer's bound can be approached quantitatively with an approaching rate of 79.3% via quantum spin tunneling. An optically-manipulated spin-encoded quantum computer is designed, in which energy bound near $k_BT$ to erase a spin qubit is theoretically sensible and experimentally verified. This work may represent the last piece of the puzzle in quantum Landauer erasure in terms of a single spin being the smallest information carrier.

## I. Introduction

Quantum computing expressed by unitary operation is notably reversible whereas the projective initialization needed to initialize the system in an entangled state and the projective measurement needed to recover classical information from the computation are not. Landauer's bound [1] limits these irreversible operations so that the increased number of computations per joule of energy dissipated will come to a halt around 2050 [2][3].

Landauer's bound was proposed in 1961, when Landauer argued that information is physical and that the erasure of a bit of classical information requires a minimum energy of $\Delta E = k_B T \ln 2$ where $k_B$ is the Boltzmann constant and $T$ is the temperature of the system. Profoundly, Landauer's principle defined the ultimate physical limit of computations [1].

In March 2012, Landauer's bound was experimentally verified by Bérut et al. in a single silica glass bead (2 $\mu m$ diameter) as a Brownian particle. The particle was trapped in a double-well potential. The mean dissipated heat was observed to saturate at the bound in the limit of long erasure cycles [4]. In June 2012, Alexei et al. reported the first experimental test of Landauer's principle in logically reversible operations, in which they measured energy dissipations much less than Landauer's bound (at the sub-$k_BT$ level) whereas irreversible operations dissipate much more than Landauer's bound [5].

In 2014, Jun et al. verified the bound in a fluorescent particle (200 $nm$). They demonstrated using small particles in traps and reducing the exerted work to the Landauer limit during erasure [6].

In 2016, Hong et al. extended the principle to orientation-encoded information and measured an energy dissipation of 4.2 zeptojoules in a single-domain nanomagnet (comprising more than $10^4$ spins). They used a laser probe to measure the energy dissipation when a bit was flipped from off to on [7].

In May 2018, a team led by Feng reported a single-atom demonstration of Landauer's principle in a fully quantum system [Fig.1(a)], in which a trapped ultracold $^{40}Ca^+$ ion was used as an atom qubit (comprised of its two internal states) [8]. The erasure procedure was completed with the heat reservoir (the ion's own vibrational modes) and the work involved was measured [8][9]. In June 2018, Gaudenzi et al. also extended Landauer's principle to the quantum realm in a collective $S_z = \pm 10$ ($20\,\mu_B$) giant spin at 1 $K$, with a superconducting quantum interference device (SQUID) [10].

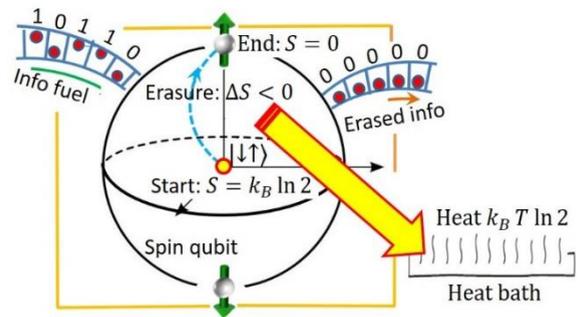

(a) A quantum computer as an information "burning" engine;

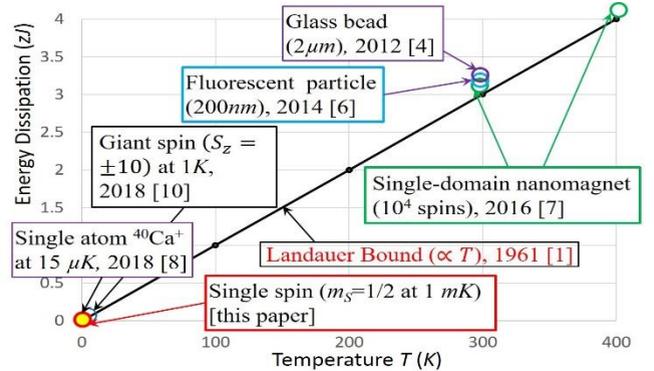

(b) Various information carriers.

**FIGURE 1.** (a) A quantum computer acting on few spin qubits, whose state can be represented by a point on the so-called Bloch sphere. At the start of an erasure, the spin is equally likely to be in either of the up/down states (corresponding to the center of the Bloch sphere) and thus has a maximal entropy $S = k_B \ln 2$. The qubit ends up in a pure quantum state (a point on the Bloch sphere's surface), in which it has a zero entropy $S = 0$ [9]. This information erasure is an irreversible manipulation of the created information with $S < 0$, i.e., the "Maxwell demon" [4] or the observer that "created" the information loses the ability to extract work from the system if the information is already "burnt". This energy bound is achievable even if a computation is carried out by a complex quantum circuit with many individual unitary gates [13]. (b) Various experimental verifications of Landauer's bound in different information carriers at their respective operating temperatures. This study on a single spin may represent the last piece of the puzzle in quantum Landauer erasure.

In March 2020, Saira et al. measured Landauer's bound at 500 $mK$ [11]. In June 2020, Çetiner et al. measured Landauer's bound in ion channels, which are smaller than the florescence molecules but larger than the spins [12].

In March 2021, Holtzman et al. showed that Landauer's bound is enforced by the contraction of the physical system's phase-space volume during the bit erasure and then suggested that, if the energy of the system is precisely known, it is possible to implement an erasable bit with no thermodynamic cost in a Hamiltonian memory [13]. However, they also pointed out that their proposal is of a purely theoretical nature and any uncertainty in the energy (i.e., the knowledge of the system's energy is limited in any realistic situation) results back in Landauer's bound [13]. In April 2021, Chiribella et al. found that even a logically reversible quantum operation (running on a physical processor operating on different energy levels) requires energy and quantified the upper and lower bounds [14]. Their bounds are present even if the evolution is entirely reversible [14]. Remarkably, their bounds can be compared quantitatively with the classical Landauer bound, which is present when the evolution is irreversible [14]. In November 2021, Georgescu reviewed 60 years of Landauer's principle and summarized that this principle imposes a fundamental energy bound on both irreversible bit operations in classical systems (which is the traditional domain of Landauer's principle) and even the reversible operations of quantum computation in spite of the distinction between these two operations [15].

Here, Landauer's bound will be studied in a single spin as the smallest information carrier among various ones, as shown in Fig.1(b). In the era of quantum computing, one naturally wonders if there is a way to ultimately approach the bound considering that quantum and classical bits are fundamentally different [9]. We will attempt to answer this question in this article.

## II. Position-Encoded Information

The statistical-mechanical formula for the free energy $F$ is: $F = -k_B T \ln Z$, where $Z$ is the partition function [16]. In one-dimensional Brownian motion, the position-encoded system [a solid particle as an information carrier trapped in a chamber with impenetrable walls as shown in Fig.2(a)] can be approximated as one in internal thermodynamic equilibrium at each given value of the coordinate $x$ of the particle. The subsystem formula is: $F(x) = -k_B T \ln Z(x)$, where $F(x)$ is the subsystem free energy at $x$, and $Z(x)$ is obtained by summation of the microstates at $x$ [16][17].

For a bit of position-encoded (classical) information in Fig.2(a), the information carrier for "random data" is equally likely to be in the $L$ or $R$ chamber, i.e. the probabilities are $P(L)=P(R)=1/2$. After erasure, the carrier is assuredly reset to a fixed reference state (the $L$ chamber in this case): $P(L)=1; P(R)=0$.

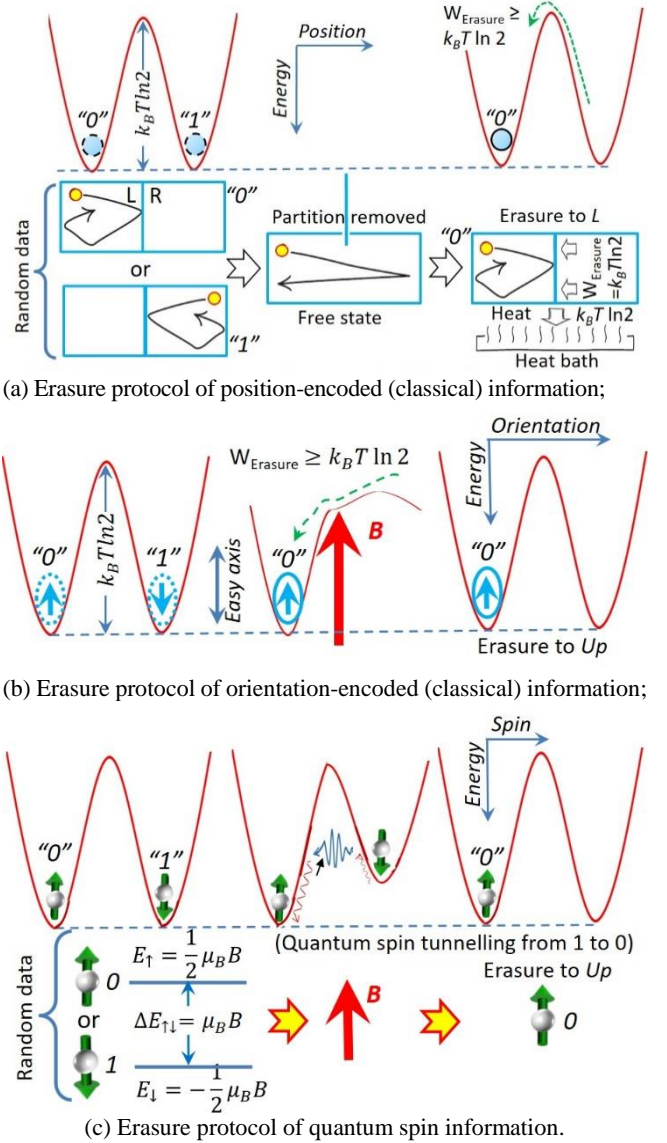

(a) Erasure protocol of position-encoded (classical) information;

(b) Erasure protocol of orientation-encoded (classical) information;

(c) Erasure protocol of quantum spin information.

**FIGURE 2.** Comparison of the three erasure protocols. For a bit of position-encoded (classical) information (e.g., a silica bead [4], and a fluorescent particle [6]) in (a), the erasure ($L$) state is reached from the random data state via the free state (the carrier can move freely between the two chambers) by removing the partition. As an isothermal contraction, the erasure creates $k_BT ln2$ (Landauer's bound) by introducing a frictionless piston and pushing it towards the $L$ direction. For a bit of orientation-encoded (classical) information (e.g., a single-domain nanomagnet comprising more than $10^4$ spins [7] and being large enough to be treated as classical [10]) in (b), the erasure ($Up$) state is reached from the random data state by applying a magnetic field $B$ along z to overcome the barrier $k_BT ln2$ (this field also tilts the potential). For a bit of quantum spin information (e.g., a $S_z=\pm10$ giant spin [10] and a single spin in this article) in (c), the erasure ($Up$) state is reached from the random data state by applying a small magnetic field $B$. In (c), the position of a wavefunction (in blue) represents the (lower-lying) quantum energy level in contrast to the classical double-well potential landscape (in red) that is needed for the Landauer erasure.

The work to push the information carrier (with two possible positions) to the desired half ($L$) is:

$$W \geq F(x) = k_B T \ln 2, \quad (1)$$

where $Z(x) = \frac{1}{2}$ since the information carrier has only two possible positions in the chamber.

Interestingly, the above operation could be performed by a "Maxwell's demon" that consumes energy to observe the position of the carrier and insert the partition, where the consumed energy still equals the work exerted for erasure.

### III. Orientation-Encoded Information

A single-domain nanomagnet, comprising more than $10^4$ spins [7] and being large enough to be treated as classical [10], was used to represent a bit of datum by encoding its (magnetization) orientation, as shown in Fig.2(b). Due to thermal agitation, the orientation ($x$) of a magnetic moment fluctuates and may therefore take an arbitrary direction. The probability [that is proportional to $Z(x)$] to find $x$ at thermal equilibrium can be deduced from Eq.1. Hence, we have:

$$F(x) = -k_B T \ln Z(x) = k_B T \ln 2, \quad (2)$$

where $Z(x) = \frac{1}{2}$ since the direction of a magnetic moment is either "up" or "down" along the easy axis. The two possible orientations of a magnetic moment are analogous to the two possible positions of a Brownian particle in the position-encoded information system. In other words, these two information systems share the same thermodynamics.

As a quantum computing paradigm, a single or giant spin can be used as a bit of quantum spin information by encoding its spin orientation, as shown in Fig.2(c). The spin angular momentum is quantized with only two possible $z$-components. decreases. At efficiently low temperatures, direct tunneling via the ground state becomes relevant and often provides the dominant spin relaxation channel [10]. As illustrated in Fig.2(c), quantum spin tunneling through the barrier from "1" to "0" is combined with excitation absorbing resonant phonons to reach the (tunneling) state and de-excitation emitting a phonon to the ground state [10].

The energy of flipping a spin undergoing a magnetic field $B$ is:

$$\Delta E_{\uparrow\downarrow} = \vec{\mu}_B \cdot \vec{B} = \vec{\mu}_B \cdot B\hat{z} = \gamma \hat{S}_z B = \gamma B \frac{\hbar}{2}(|\uparrow\rangle\langle\uparrow| - |\downarrow\rangle\langle\downarrow|), \quad (3)$$

where $\mu_B$ is the Bohr magneton, $\gamma$ is the gyromagnetic ratio of an isolated electron, $\hat{S}_z = \frac{\hbar}{2}\begin{bmatrix} 1 & 0 \\ 0 & -1 \end{bmatrix}$ is the quantum-mechanical operator associated with spin-$\frac{1}{2}$ observable in the $z$ axis, and $\hbar$ is the reduced Planck constant.

Superficially, this new energy bound of flipping a spin is decoupled from the environmental temperature $T$, but, taking the giant spin as an example, (phonon-mediated/assisted) quantum spin tunneling is still coupled to the 'surrounding world', including the environmental temperature $T$ [10]. Namely, the spin relaxation time, $\tau_{rel}$ approximately follows Arrhenius's law: $\tau_{rel} = \tau_0 \exp(\frac{U}{k_B T})$, where $\tau_0 = 10^{-8} s$, $U$ is the activation energy determined by the tunneling channel and $\tau_{rel} \geq 100\ s$ as $T$ decreases to $1\ K$ [10].

### IV. Experiment of Spin-Spin Magnetic Interaction

Extremely weak magnetic interactions between the two ground-state spin-1/2 ($1\ \mu_B$) valence electrons of two $^{88}$Sr$^+$ ions across a separation ($d = 2.18 \sim 2.76\ \mu m$) were reportedly measured as shown in Fig.3 [18]. The two ions were trapped in a linear *rf* Paul trap with a radial trap frequency ($\Gamma = 2\pi \times 2.5$ MHz) and laser-cooled to $1\ mK$ [18][19][20]. The measurement takes full advantage of the quantum lock-in method [19] to spectrally separate weak signals from noise.

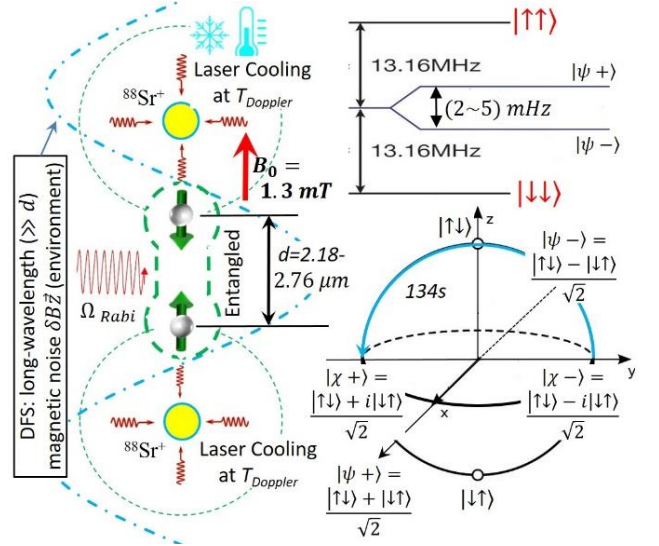

**FIGURE 3.** Extremely weak magnetic interaction between the two ground-state spin-1/2 ($1\ \mu_B$) valence electrons of two $^{88}$Sr+ ions was reportedly measured [18]. This magnetic interaction can then impose a change in their orientation. The two ions were laser-cooled to their ground state and entangled across a separation ($d = 2.18 \sim 2.76\ \mu m$). An ion will absorb more photons if they move towards the light source and the net result is a reduced speed of the ion, which is equivalent to cooling the ion since the temperature is a measure of the random internal kinetic energy. An external magnetic field $B_0 = 1.3 \times 10^{-3}\ T$ sets the spin quantization axis and lifts the degeneracy between the probe states by $f_0 = 13.16\ MHz$. The measured energy splitting ($2 \times 13.16\ MHz$) between $|\uparrow\uparrow\rangle$ and $|\downarrow\downarrow\rangle$ in this experiment can be fully used to verify the calculated energy $\Delta E_{\uparrow\downarrow}$ of (irreversibly) flipping a spin in Eq.5. A redraw courtesy of Shlomi Kotler (the Hebrew University of Jerusalem).

In this experiment, it was found that the spin–spin magnetic interaction obeys the inverse-cube law and spin entanglement was observed [18]. As the smallest magnet (the Bohr magneton), a spin ($\vec{\mu}_B$) applies a magnetic field to another spin. While the two spins are aligned along the line connecting the two ions [18], the strength of the magnetic field is:

$$B_{spin-spin} = \frac{\mu_0}{4\pi} \frac{2\mu_B}{d^3} = (0.88 \sim 1.79) \times 10^{-13} \, T, \quad (4)$$

where $\mu_0$ is the vacuum permeability constant. This equivalent magnetic field $B_{spin-spin}$ is ten orders of magnitude smaller than the actual magnetic field $B_0 = 1.3 \times 10^{-3}$ T.

According to Eq.3, we should use $B_0$ to compute the energy of flipping a spin as below:

$$\Delta E_{\uparrow\downarrow} = \mu_B B_0 = 9.274 \times 10^{-24} \, J/T \times 1.3 \times 10^{-3} \, T = 1.21 \times 10^{-26} \, J. \quad (5)$$

This energy converted to $\frac{\Delta E_{\uparrow\downarrow}}{h} = \frac{1.21 \times 10^{-26} \, J}{6.63 \times 10^{-34} \, J \cdot s} = 18.3 \, MHz$ is reasonably close to the measured frequency (13.61 MHz) in the spin-spin magnetic interaction experiment [18].

A fault-tolerant quantum computer with imperfect quantum logic gates in practice needs to perform long computations without succumbing to some inevitable errors and noise, which raises a good concern in reliability or error probability. We stress that a single spin can be switched reliably (with a typical detection fidelity of 98% in the presence of magnetic noise that is six orders of magnitude greater than the applied magnetic field) [18]. The spin evolution was restricted to a decoherence-free subspace (DFS) that is immune to collective magnetic field noise [18] since the two spins "see" the same (time-dependent) magnetic noise from the environment, whose wavelength is much larger than the separation $d$ (Fig.3).

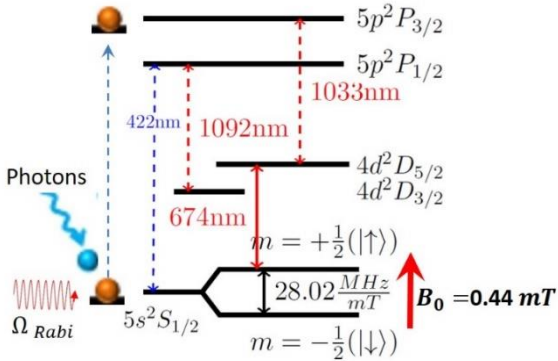

**FIGURE 4.** A single $^{88}Sr^+$ ion's degeneracy between the probe states was lifted (by $f_0 = 12.34 \, MHz$) by applying a magnetic field $B_0 = 0.44 \, mT$. The corresponding gyromagnetic ratio is $\gamma = \frac{12.34 \, MHz}{0.44 \, mT} = 28.02 \, \frac{MHz}{mT}$. Spin rotations were performed by pulsing a resonant oscillating magnetic field, perpendicular to the quantization axis (set by $B_0$), resulting in a Rabi frequency $\Omega_{Rabi} = 65.8 \, kHz$ [18].

As a popular technique to raise (or "pump") electrons from a lower energy level in an atom or molecule to a higher one, optical pumping was also used here to pump electrons bound within the ions to a well-defined quantum state $|\uparrow\downarrow\rangle$ or $|\downarrow\uparrow\rangle$, as shown in Fig.4. The frequency and polarization of the pump laser determined the sublevel in which the electron was oriented. This experiment displayed the ability of coherent electromagnetic radiation (having wavelengths below one millimeter) to effectively pump and unpump these electrons. The infrared 1092 nm and 1033 nm lasers acted as repump lasers [18]. Generation of entangled Bell states of the form $|\psi\pm\rangle = (|\uparrow\downarrow\rangle \pm |\downarrow\uparrow\rangle)/\sqrt{2}$ was done using a Sörenson-Mölmer entangling gate [18]. A pure quantum state represented by the Bloch vector can be located by measuring its projection on an equal superposition, e.g. the $|\chi\pm\rangle = (|\uparrow\downarrow\rangle \pm i|\downarrow\uparrow\rangle)/\sqrt{2}$ basis (i.e., y) if rotating it around $x$, via a parity observable. This collective rotation does not change the relative orientation of the spins, leaving the spin–spin interaction invariant [18]. The parity observable measures the coherence between $|\uparrow\downarrow\rangle$ and $|\downarrow\uparrow\rangle$: it is +1 if the spins are aligned and −1 if they are anti-aligned.

In magneto-optical traps (MOTs), the actual temperature is $(10 \sim 30) T_{Doppler}$ [21]. The minimum Doppler temperature is:

$$T_{Doppler} = \frac{\hbar\Gamma}{2k_B} = \frac{1.05 \times 10^{-34} J \cdot s \times 2\pi \times 2.5 \times 10^6 s^{-1}}{2 \times 1.38 \times 10^{-23} J \cdot K^{-1}} = 5.07 \times 10^{-5} K, (6)$$

where $\Gamma$ is broad natural linewidth (measured in radians per second), hence the calculated temperature is $T = (10 \sim 30) \times 5.07 \times 10^{-5} K = (0.51 \sim 1.52) mK$, which agrees reasonably with the measured temperature of 1 $mK$. Landauer's bound can be expressed by $k_B T \ln 2 = k_B T \ln 2 = 9.6 \times 10^{-27} J$ at 1 $mK$, which is $10^{-5}$ times Landauer's bound ($3 \times 10^{-21} J$) at room temperature (300 $K$) as it is proportional to the temperature.

Noticeably, according to Eq.5, the input energy ($1.21 \times 10^{-26} J$) to erase a spin quantum datum is very close to Landauer's bound ($9.6 \times 10^{-27} J$) at 1 $mK$. This verification simply makes full use of the measured data from this spin-spin experiment backdated to 2014 [18], whose authors wrote to us "It's exciting to hear that our work is useful in new areas of research that we were not aware of when doing the experiment." after we shared this manuscript with them.

Although the spin-spin experiment [18] is in a completely different (magnetic-interaction with the inverse-cube law) context, it equivalently includes a complete erasure protocol with $S < 0$ and gives the measurement of the work involved, as shown in Fig.5. This equivalence is based on $|\downarrow\uparrow\rangle = \frac{1}{\sqrt{2}}(|\psi+\rangle - |\psi-\rangle) = \frac{1}{\sqrt{2}}\left(\left[\frac{|\uparrow\downarrow\rangle + |\downarrow\uparrow\rangle}{\sqrt{2}}\right] - \left[\frac{|\uparrow\downarrow\rangle - |\downarrow\uparrow\rangle}{\sqrt{2}}\right]\right)$. This complete erasure protocol has all the necessary steps as defined in [8]: at the start step of erasure, a circularly on-resonant 422 nm laser was used to cyclically pump the two electrons bound within the two ions to a maximally mixed quantum state (see Fig.4 for details): the spin is equally likely to be in either of the up/down states (corresponding to the center of the Bloch sphere) and thus has a maximal entropy $S = k_B \ln 2$; at the mediate step of erasure, the (optically) created qubit is then erased by the external magnetic field; at the end step of erasure, the qubit ends up in a (ground) quantum state $|\uparrow\rangle$ (a point on the Bloch sphere's surface), in which it has a zero entropy $S=0$. Most importantly, the decreased entropy ($S<0$) ensures the irreversibility required in a Landauer erasure process.

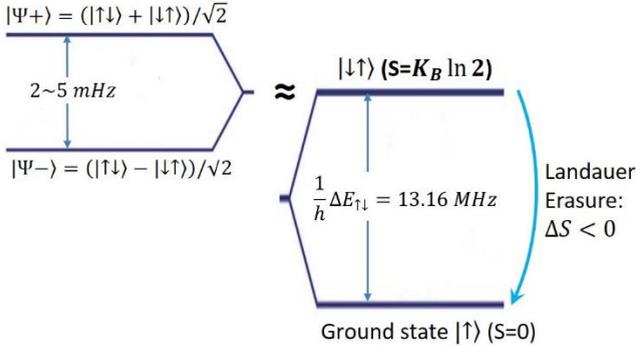

**FIGURE 5.** Two $^{88}Sr^+$ ions' energy diagram. Although the spin-spin experiment [18] is in a completely different (magnetic-interaction) context, it equivalently includes a complete erasure protocol and gives the measurement of the work involved. $|\downarrow\uparrow\rangle = \frac{1}{\sqrt{2}}(|\psi+\rangle - |\psi-\rangle) = \frac{1}{\sqrt{2}}\left(\left[\frac{|\uparrow\downarrow\rangle + |\downarrow\uparrow\rangle}{\sqrt{2}}\right] - \left[\frac{|\uparrow\downarrow\rangle - |\downarrow\uparrow\rangle}{\sqrt{2}}\right]\right)$. **In such an (irreversible) information erasure ($|\downarrow\uparrow\rangle \to |\uparrow\rangle$), the work one needs to do in order to reset a bit register irrespective of its initial state has to compensate for the illustrated entropy drop $\Delta S$ [8][9].**

To verify the completeness (start/erasure/end) of the erasure protocol, one may compare the spin-spin experiment [18] and the single-atom demonstration that completed the erasure protocol to support quantum Landauer principle [8]. Essentially differently, in this study of a single spin as the smallest information carrier, we attempt to verify that Landauer's bound holds at a size of a single spin.

To identify the dominant factors in our problem, we rewrote (the spin part of) the two-ion Hamiltonian in the spin-spin experiment [18] as:

$$H = \underbrace{0.5\hbar(\omega_{A,1}\sigma_{z,1} + \omega_{A,2}\sigma_{z,2})}_{Magnetic\ field\ B_0=1.3\ mT\ (MHz)}$$
$$+\underbrace{2\hbar\zeta\sigma_{z,1}\sigma_{z,2}}_{Spin-spin\ (mHz)} \underbrace{-\hbar\zeta(\sigma_{x,1}\sigma_{x,2} + \sigma_{y,1}\sigma_{y,2})}_{Rabi\ flopping\ (kHz)}. \quad (7)$$

Here $\sigma_{j,i}$ is the $j \in \{x,y,z\}$ Pauli spin operator of the $i$th spin, within which $\sigma_{z,1}\sigma_{z,2}$ does not cause any spin-flips and acts as a phase gate in quantum computing whereas $\sigma_{x,1}\sigma_{x,2}$ and $\sigma_{y,1}\sigma_{y,2}$ lead to Rabi flopping of $|\uparrow\downarrow\rangle \leftrightarrow |\downarrow\uparrow\rangle$; $\omega_{A,i} = 2\mu_B B_i/2\hbar$, where $B_i$ is the external magnetic field. The spin–spin interaction strength is $\zeta = \mu_0\mu_B^2/4\pi\hbar d^3$. The first term on the right-hand side of Eq.7 describes the Zeeman shift of the spins' energy due to the external magnetic field $B_0 = 1.3\ mT$, which is equivalent to $MHz$ in the spin Larmor frequency $\omega_{A,i}$ ($i=1,2$) [18] that characterizes the precession of a transverse magnetization about a static magnetic field. The second term describes the spin–spin magnetic interaction, which is equivalent to 2-5 $mHz$ [18]. The third term results in a collective spin flip, in which spin rotation is performed by pulsing a resonant oscillating magnetic field, resulting in a Rabi frequency in $kHz$ [18]. The spin Larmor frequency in the first term on the $MHz$ order whereas the Rabi frequency in the third term is on the $kHz$ order, the former is dominant in terms of calculating the work of irreversibly erasing a spin qubit. It is the first term (13.16 $MHz$) that is at the focus of our study.

An analytical model to explain the above experimental verifications follows.

### V. Spinor Wavefunction of An Isolated Electron
The Schrödinger-Pauli equation for an isolated electron [the smallest magnet being an information carrier shown in Fig.6(a)] is:

$$i\hbar \frac{d|\Psi\rangle}{dt} = \hat{H}|\Psi\rangle, \quad (8)$$

where the spinor wavefunction is $|\Psi(t)\rangle = C^+(t)|\uparrow\rangle + C^-(t)|\downarrow\rangle$, and the Hamiltonian is $\hat{H} = -\gamma B \frac{\hbar}{2}(|\uparrow\rangle\langle\uparrow| - |\downarrow\rangle\langle\downarrow|)$ according to Eq.3. Substitutions into Eq.7 give:

$$i\hbar(\dot{C}^+|\uparrow\rangle + \dot{C}^-|\downarrow\rangle) = -\gamma B \frac{\hbar}{2}(|\uparrow\rangle\langle\uparrow| - |\downarrow\rangle\langle\downarrow|)(C^+|\uparrow\rangle + C^-|\downarrow\rangle)$$
$$= -\gamma B \frac{\hbar}{2}(C^+|\uparrow\rangle - C^-|\downarrow\rangle), \quad (9)$$

$$\begin{bmatrix}\dot{C}^+ \\ \dot{C}^-\end{bmatrix} = \frac{i}{2}\gamma B \begin{bmatrix}1 & 0 \\ 0 & -1\end{bmatrix}\begin{bmatrix}C^+ \\ C^-\end{bmatrix} = \frac{i}{2}\gamma B \begin{bmatrix}C^+ \\ -C^-\end{bmatrix}. \quad (10)$$

The WKB (Wentzel–Kramers–Brillouin) approximation rewrites the (complex-valued) spinor wavefunction as:

$$|\Psi(t)\rangle = \begin{bmatrix}C^+(t) \\ C^-(t)\end{bmatrix} = \begin{bmatrix}C^+(0)e^{\Phi(t)} \\ C^-(0)e^{-\Phi(t)}\end{bmatrix}. \quad (11)$$

The time evolution takes place in the presence of an external magnetic field $B_0$. To overcome the thermal perturbation (Landauer's bound), we should have $\Delta E_{\uparrow\downarrow} = \mu_B B \approx \mu_B B_0 \geq k_B T \ln 2$. Without losing generality, $\Delta E_{\uparrow\downarrow}(t)$ is assumed as a positive constant $E$ during $-t_E/2 \leq t \leq t_E/2$. Then, we obtain:

$$\Phi\left(t = \frac{t_E}{2}\right) = i\frac{1}{\hbar}\int_{-\infty}^{t}\left(-\gamma B\frac{\hbar}{2}\right)dt\bigg|_{t=t_E/2}$$
$$= i\frac{1}{\hbar}\int_{-\infty}^{t}\Delta E_{\uparrow\downarrow}(t)dt\bigg|_{t=t_E/2}$$
$$= i\frac{1}{\hbar}Et_E. \quad (12)$$

Then, Eq.11 simplifies to:

$$|\Psi(t = t_E/2)\rangle = \begin{bmatrix}C^+(0)e^{\Phi(t)} \\ C^-(0)e^{-\Phi(t)}\end{bmatrix} = \begin{bmatrix}C^+(0)e^{i\frac{1}{\hbar}Et_E} \\ C^-(0)e^{-i\frac{1}{\hbar}Et_E}\end{bmatrix}. \quad (13)$$

Eq.13 shows that, behaving like a free and oscillating wave, the single spin with less energy tunnels through the energy hill and appears on the other side with a probability $|\Psi|^2$ to complete a reversal in the spin-spin magnetic interaction experiment [18]. A similar (quantum spin tunneling) phenomenon was observed in a collective $S_z = \pm 10$ ($20\,\mu_B$) giant spin [10].

In Eq.13, it is ($Et_E$) that defines the wavefunction. In other words, it is the energy-time product, rather than any of these two parameters ($E, t_E$) individually, that determines the behavior of the spin datum. We see that the probability of tunneling is affected more by ($Et_E$) than by $C^{+/-}(0)$. It seems that the quantum erasure differs dramatically from its classical counterpart.

In dissipative dynamics, erasing a bit of information requires probability concentration in phase space, which leads to Landauer's bound. In Hamiltonian dynamics, it is possible to take a particle from say the left well to the right one at zero cost (or as low as you want) [13]. Therefore, the problem in a Hamiltonian memory may be that, at the same time, the particle in the right well goes to the left well (or somewhere else - in any case it does not stay on the same well). Fortunately, the tunneling in the spin-spin magnetic interaction experiment [18] is irreversible since the energy is input by applying a magnetic field that only favors and flips a spin with the opposite direction and a single spin can be switched reliably with a typical detection fidelity of 98% [18]. As mentioned above, a similar phenomenon (the spins can tunnel to the opposite side of the potential barrier, thus leading to an effectively lower activation energy for the spin reversal) was also observed in a collective $S_z = \pm 10$ ($20\,\mu_B$) giant spin [10]. That is, the erasure of the spin datum in the giant spin experiment [10] and the spin-spin experiment [18] is not pure Hamiltonian dynamics and the probability concentration in phase space can still be seen.

## VI. Using Heisenberg's Principle to Define Information

To further interchange information with energy over time, we used Heisenberg's time-energy uncertainty relation (TEUR) in 1927 [23] to define information, as illustrated in Fig.6(b). From a measuring perspective, one bit of information is the smallest error in physical measurement. A bit of information is quantitatively defined as follows:

$$1\,(bit) = \frac{2}{\hbar}\Delta E \Delta t, \quad (14)$$

where we embraced a new interpretation of the TEUR: a quantum state with spread in energy $\Delta E$ takes time at least $\Delta t$ to evolve to an orthogonal (distinguishable) state [24].

Note that the above mentioned "one bit of information as the smallest error in physical measurement" should not be interpreted as "the smallest error one makes is one bit when mapping the measured analog value to a discrete sequence of digits". Here one bit is physically a quantum as the minimum amount of a conjugate pair of observables (energy/time) involved in an interaction. According to Heisenberg's TEUR, this amount corresponds to Planck's reduced constant ($\hbar = 1.054571817 \times 10^{-34}\,J \cdot s$) that defines the quantum nature of energy and relates the energy of a photon to its frequency.

Ergo, this new definition of information reflects the essence of quantum physics: the magnitude of the physical property can take on only discrete values consisting of integer multiples of one quantum (a multiple of Planck's reduced constant). Also note that $2\Delta E \Delta t/\hbar$ is unitless, which does not violate the definition of information in units.

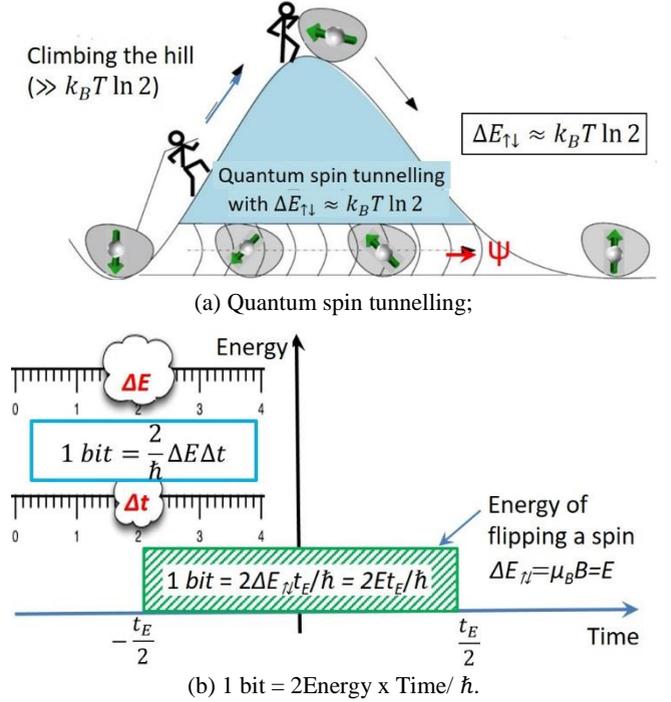

(a) Quantum spin tunnelling;

(b) 1 bit = 2Energy x Time/ $\hbar$.

**FIGURE 6. (a) Quantum spin tunneling penetrates the thermal energy barrier (much larger than Landauer's bound) and provides a "shortcut" for spin reversal, which is different from classical information manipulations. The cost in erasing a bit does not come from "climbing a barrier", but rather from compressing phase space with dissipative dynamics. (b) Heisenberg's time-energy uncertainty relation (TEUR) [23][24] is used to define information quantitatively from a measuring perspective: the smallest error in measurement is 1 bit. The higher the input energy is, the shorter the time is needed to write/erase a bit of information and vice versa. This new definition of information is an important part of our theory in terms of the energy time product being a constant, as vividly illustrated here. That is, energy close to $k_B T$ to erase a spin qubit is theoretically sensible and experimentally verified due to this unchanged product (the shaded areas). Our new definition of information based on Heisenberg's principle allows us to determine the trade-off between energy and speed of manipulating a spin qubit.**

This energy-time product is ultimate for a bit of information no matter what kind of information carrier (a bead, an atom, an ion, a nanomagnet, a giant spin, a single spin, or a photon) is used and what mechanism [classical physics (electrical, magnetic, optical, chemical or even mechanical), or quantum physics] is used to encode/manipulate it.

If Landauer's bound at room temperature is used, the time needed to write/erase a bit of information (that is physically equivalent to the duration of the energy measurement in the

TEUR [24] since energy is consumed throughout the write/erase protocol) is:

$$\Delta t = \frac{\hbar}{2\Delta E} = \frac{1.05 \times 10^{-34} \, J \cdot s}{2 \times 3 \times 10^{-21} \, J} = 1.75 \times 10^{-14} \, s. \quad (15)$$

This calculation result agrees reasonably with the Brillouin's principle [25].

Historically, more than one definition of information existed [26][27][28], which implies that information can be studied from different angles and its definition may not be unique.

The above analysis clearly shows that Landauer's bound can be approached quantitatively in such a single spin in terms of the bound being defined as the smallest amount of the energy used to erase a bit of information.

## VII. Conclusion & Discussions

We depict an optically-manipulated spin-encoded quantum computer (Fig.1) that can approach Landauer's bound. This study and may represent the last piece of the puzzle in quantum Landauer erasure.

Today's few-qubit quantum computers require large cooling machinery external to the actual quantum processors whereasthe fundamental energy requirement as given by Eq.5 merely represents a minor part of the overall energy bill. However, with the progress of the quantum technology, the cooling energy is likely to scale less than linearly with the number of qubits, hence its proportion may become less dominant [14]. Nonetheless, such a spin-encoded quantum computer may be slow although it can operate at the ultimate (energy) limit to computation set by physics, as mentioned in Section VI.

Landauer's bound is widely accepted as one of the fundamental limits in computer science and physics, but it has still been challenged for using circular reasoning and faulty assumptions [29]. In 2000, Shenker argued that Landauer's dissipation thesis (logically irreversible operations are dissipative by $k_B \ln 2$ per bit of lost information) is plainly wrong since logical irreversibility has nothing to do with dissipation [30]. In 2003, Bennett suggested that a no-erasure demon is subject to an extended form of Landauer's principle to refute Shenker's argument and claimed that, although in a sense it is indeed a straightforward consequence or restatement of the second law of thermodynamics, it still has considerable pedagogic and explanatory power [31]. In 2005, Norton pointed out that, due to the illicit formation, Bennett's extension in order to exorcise the no-erasure demon failed [32]. In 2007, Ladyman et al. defended the qualitative form of Landauer's Principle, and clarified its quantitative consequences (assuming the second law of thermodynamics) [33]. In 2008, Sagawa and Ueda showed that Landauer's principle is a consequence of the second law of thermodynamics with discrete quantum feedback control [34]. In 2009, Cao and Feito illustrated some consequences by computing the entropy reduction in feedback controlled systems [35]. In 2011, Norton showed that the previous proofs selectively neglect thermal fluctuations that may fatally disrupt their intended operation [36]. In 2019, Jordan and Manikandan disagreed with Norton and found the principle to be easily derivable from basic principles of thermodynamics and statistical physics [37]. In 2019, Norton argued that Jordan and Manikandan were mistaken with their saying (dissipation is only necessitated when logically irreversible processes are required) since the existence of thermal fluctuations and the high thermodynamic cost of suppressing them are still unavoidable [38].

In light of the above research, we will further investigate those direct/indirect proofs [36] of Landauer's principle to see whether it is just a direct consequence or restatement of the second law of thermodynamics (the information erasure results in a decreased entropy). This investigation is important and necessary no matter whether we still want to regard Landauer's principle as fundamental as the second law of thermodynamics or not. We will also study whether it is possible to implement an erasable bit without thermodynamic cost by compressing phase space with dissipative dynamics [39][40][41].

In spite of plenty of mysteries with Landauer's bound, we may have to presume its demise based on the concerns (something is fundamentally awry in the literature based on unsound, incoherent foundations, principles, methods and/or frameworks) expressed by those researchers [29][30][32][34][35][36][38]. As well as having significant practical importance, understanding the fundamental limits on what we can achieve with our computing machines [24] is nothing less than understanding the limits of the world in which we live and preparing for revolutions, such as energy-efficient quantum computing paradigms [42].

**Acknowledgements**
We thank Dr. Shlomi Kotler (the Hebrew University of Jerusalem) for discussions on his magnetic spin-spin interaction experiment and his kind permission for us to redraw the experimental setup. We also thank Dr. Sai Vinjanampathy (Indian Institute of Technology Bombay) for commenting the first draft of this paper. This research was partially funded by an EC grant, PIIFGA2012332059, Marie Curie Fellow: Prof. Leon Chua (UC Berkeley), Scientist-in-charge: Prof. Frank Wang (University of Kent).

**Author Contributions**
Frank Wang conceived the research idea, analyzed all the experiments, developed the new theory to explain the experimental verifications, and wrote the manuscript.

**Data availability**
All data generated and analysed during this study are included in this published article.